\documentclass[]{aastex631}

\shorttitle{A Model RRNet for Spectral Information Exploitation}
\shortauthors{Xiong et al.}

\graphicspath{{./}{images/}}

\usepackage{CJKutf8}

\begin{document}
\begin{CJK}{UTF8}{gbsn}

\title{A Model RRNet for Spectral Information Exploitation and LAMOST Medium-resolution Spectrum Parameter Estimation}

\author{Shengchun Xiong}
\affiliation{School of Computer Science, South China Normal University, 510631 Guangzhou, People’s Republic of China}

\author[0000-0003-3182-6959]{Xiangru Li}
\affiliation{School of Computer Science, South China Normal University, 510631 Guangzhou, People’s Republic of China}

\author{Caixiu Liao}
\affiliation{School of Mathematical Sciences, South China Normal University, 510631 Guangzhou, People’s Republic of China}

\begin{abstract}
This work proposes a Residual Recurrent Neural Network (RRNet) for synthetically extracting spectral information, and estimating stellar atmospheric parameters together with 15 chemical element abundances for medium-resolution spectra from Large Sky Area Multi-Object Fiber Spectroscopic Telescope (LAMOST).
The RRNet consists of two fundamental modules: a residual module and a recurrent module.
The residual module extracts spectral features based on the longitudinally driving power from parameters, while the recurrent module recovers spectral information and restrains the negative influences from noises based on Cross-band Belief Enhancement.
RRNet is trained by the spectra from common stars between LAMOST DR7 and APOGEE-\textit{Payne} catalog.
The 17 stellar parameters and their uncertainties for 2.37 million medium-resolution spectra from LAMOST DR7 are predicted.
For spectra with $\rm{S/N \geq 10}$, the precision of estimations ($T_{\rm eff}$ and $\log \, g$) are 88 K and 0.13 dex respectively, elements C, Mg, Al, Si, Ca, Fe, Ni are 0.05 dex to 0.08 dex, and N, O, S, K, Ti, Cr, Mn are 0.09 dex to 0.14 dex, while that of Cu is 0.19 dex.
Compared with \textit{StarNet} and \textit{SPCANet}, RRNet shows higher accuracy and robustness.
In comparison to Apache Point Observatory Galactic Evolution Experiment and Galactic Archaeology with HERMES surveys, RRNet manifests good consistency within a reasonable range of bias.
Finally, this work releases a catalog for 2.37 million medium-resolution spectra from the LAMOST DR7, the source code, the trained model and the experimental data respectively for astronomical science exploration and data processing algorithm research reference.

\end{abstract}


\section{Introduction} \label{sec:intro}

With the gradual development of various large-scale spectroscopic surveys \citep[e.g.][]{steinmetz2006radial, yanny2009segue, gilmore2012gaia, de2015galah, luo2015first, majewski2017apache}, a large number of spectra are observed along with them.
The spectra of these surveys provide important data supports for astronomers to investigate fundamental astronomical problems.
The spectra contain numerous stellar information, such as effective temperature ($T_{\rm eff}$), surface gravity ($\log \, g$), and elemental abundances.
These information can be used in studying the formation and evolution of the Milky Way \citep{Frankel2018Measuring, Bland-Hawthorn2019theGALAH}.

The majority of spectroscopic surveys are equipped with pipelines for obtaining stellar parameters and element abundances.
These pipelines commonly match observed spectra with theoretical (or empirical) spectra by minimizing $\chi^2$ distance, and using the labels of the best-matched theoretical spectra as the parameter estimation from the observed spectra.
For example, LAMOST developed the stellar parameter pipeline LASP \citep{Wu2011Automatic, luo2015first} based on the ULySS package \citep{Koleva2009ULySS};
APOGEE is equipped with the stellar parameters and chemical abundance pipeline ASPCAP \citep{Garcia2016ASPCAP, Jonsson2018APOGEE} for near-infrared spectra;
GALAH applied the Spectroscopy Made Easy tool to obtain the stellar parameters \citep{Piskunov2017SME}.
However, the application of the above-mentioned methods to LAMOST spectra at medium-resolution (R $\thicksim$ 7500) or low-resolution (R $\thicksim$ 1800) faces the following challenges:
The theoretical spectra come from stellar atmosphere models which depend on overly simplified physical assumptions, resulting in some gaps between theoretical and measured spectra.
In addition, the presence of blended features in spectra makes matching the correct theoretical spectra more difficult.

With the arrival of artificial intelligence and the big data era, deep learning methods have been attempted to deal with the estimation of stellar parameters.
This kind methods usually estimate parameters by approximating the mapping relationship from observed spectra to stellar parameters \citep[e.g.][]{fabbro2018application, Bialek2020Assessing, leung2019deep, wang2020spcanet}, or from stellar parameters to observed spectra \citep[e.g.][]{ ting2019payne, Xiang2019Abundance, rui2019analysis} using neural networks.
However, most of the above models use simple fully-connected neural networks or convolutional neural networks to establish the mapping relationships.
While, it is shown that these methods are difficult to extract some deep spectral features from low signal-to-noise spectra.

To this end, we designed a Residual Recurrent Neural Network (RRNet), and established a mapping from LAMOST medium-resolution spectra to stellar atmospheric parameters and elemental abundances using deep ensembling.
The RRNet consists of two fundamental modules to comprehensively extract spectral features: the residual module and recurrent module.
The residual module is longitudinally driven by the spectrum labels in the training data set to extract features, while the recurrent module exploits spectral features and restrains the negative influence from noise based on Cross-band Belief Enhancement (CBE) between the observations on different wavelength subbands.
The CBE refers to the maximum information recovery and extraction by fusing the observations on different wavelength subbands in case of the existence of some correlations between them and some difference of disturbances on them (See Fig. \ref{fig:figure4}).
Som LAMOST spectra with APOGEE-\textit{Payne} label was extracted as reference spectra. These spectra were observed from 28,523 common stars between LAMOST DR7\footnote{\href{http://www.lamost.org/dr7/}{http://www.lamost.org/dr7/}} and APOGEE-\textit{Payne} catalog \citep{ting2019payne}. The reference spectra are used for training and testing the RRNet model.
Finally, stellar atmospheric parameters, elemental abundances and the $1\sigma$ uncertainties of the corresponding parameter estimations are derived for 2,377,510 medium-resolution spectra from LAMOST DR7 by RRNet.

This paper is organized as follows: 
Section \ref{sec:Data} introduces the APOGEE-\textit{Payne} catalog, the LAMOST DR7 medium-resolution spectra, and the associated data pre-processing methods.
Section \ref{sec:Method} describes the RRNet model and its validation experiments. 
Section \ref{sec:Results} presents the results of our study.
Summary and outlook are made in Section \ref{sec:Summary}. 
The computed catalog for 2.37 million medium-resolution spectra from the LAMOST DR7, the source code, the trained model and the experimental data are released on: \href{https://github.com/Chan-0312/RRNet}{https://github.com/Chan-0312/RRNet}.

\section{Data}\label{sec:Data}
To learn the model parameters of RRNet, a reference set is needed.
The reference dataset consists of some LAMOST DR7 medium-resolution spectra with the labels of their stellar parameters ($T_{\rm eff}$, $\log \, g$) and 15 elemental abundances ([C/H], [N/H], [O/H], [Mg/H], [Al/H], [Si/H], [S /H], [K/H], [Ca/H], [Ti/H], [Cr/H], [Mn/H], [Fe/H], [Ni/H], [Cu/H]) from the APOGEE-\textit{Payne} catalog. 
This reference set is established by cross-matching the observations between LAMOST DR7 medium-resolution spectra and the APOGEE-\textit{Payne} catalog.
The establishment of the reference set and the related pre-processing procedures are described furtherly in the following two subsections.

\subsection{Reference dataset}

LAMOST provides a large number of precious spectra for researchers.
LAMOST began its Phase II medium-resolution survey in July 2017 and released 5.6 million medium-resolution spectra in LAMOST DR7, with a total of 2.4 million spectra with a signal-to-noise ratio (S/N) greater than 10 \citep{rui2019analysis}.
During the survey, two spectra are obtained for each exposure, one is the blue part and the other is the red part, and their wavelength coverages are [4950, 5350] \text{\AA} and [6300, 6800] \text{\AA} respectively.

\cite{ting2019payne} analyzed the spectra from APOGEE DR14 based on the Kurucz theoretical model, obtained a high-precision APOGEE-\textit{Payne} catalog.
The catalog gives stellar parameters for 222,702 stars, including $T_{\rm eff}$, $\log \, g$, and 15 elemental abundances.
The coverages of the stellar atmospheric parameters in the APOGEE-\textit{Payne} catalog are $\rm [3050, 7950]$ K on $T_{\rm eff}$, $\rm [0, 5]$  dex on $\log \, g$, and $\rm [-1.45, 0.45]$ dex on $\rm [Fe/H] $.
The accuracies of the three parameters are 30 K, 0.05 dex and 0.05 dex respectively.

Following \cite{wang2020spcanet}, we cross-matched the LAMOST DR7 medium-resolution spectra with the APOGEE-\textit{Payne} catalog, obtained  161,447 LAMOST DR7 spectra from 34,372 common stars.
To ensure the reliability of the dataset, the LAMOST spectra with $\rm S/N < 10$ are eliminated from the cross-matched data set. 
In addition, some LAMOST spectra are affected by cosmic rays and other influences, which result in a large number of outliers (bad pixels) in them. 
Therefore, the spectra with more than 100 outliers or more than 30 consecutive outliers are rejected in this work.
In the APOGEE-\textit{Payne} catalog, we kept the information from the spectra with $\rm quality\_flag = good$ and $T_{\rm eff} \in \rm [4000, 6500]$ K.
Through the above processing, we finally obtained 28,523 common stars and 114,853 medium-resolution spectra of the common stars from LAMOST DR7.
These spectra are used as the reference data for our model. 

\subsection{Data pre-processing}\label{sec:Data:preprocess}

To facilitate the optimization of the model, we pre-process the LAMOST spectra as follows:

\textbf{Wavelength correction}: The Radial Velocity (RV) can broaden or narrow down the spectral lines in the observed spectrum. 
To account for this kind of broadening and narrowing down effects, a much large reference set should be used in training a machine learning parameter estimation scheme for a similar accuracy. 
To simplify the spectral parameter estimation problem based on the available reference set, we perform wavelength correction on each spectrum by shifting it to its rest frame using the RV provided by the LAMOST catalog:
\begin{equation}
\label{equ:equation1}
\lambda'= \frac{\lambda}{1 + \rm RV/ \it c},
\end{equation}
where $\lambda'$ is the corrected spectral wavelength in the rest frame, $\lambda$ is the observed spectral wavelength, RV is the radial velocity of the corresponding spectrum, and $c$ is the speed of light.
(For more discussion see appendix \ref{sec:dis_wavelength_correction}.)

\textbf{Spectral resampling}: According to the distribution of the spectral wavelength ranges from the reference set, the common part of the corrected spectral wavelength ranges are computed. 
The common wavelength ranges on blue part and red part are respectively [4968, 5328] \text{\AA}, and [6339, 6699] \text{\AA}.
On the common wavelength range, each spectrum is resampled using a linear interpolation procedure with a step 0.1 \text{\AA}.
Finally, 7200 fluxes (3600 in the blue part and 3600 in the red part) are obtained for each observed spectrum.

\textbf{Spectral normalization}: Consistent with \cite{wang2020spcanet}, each resampled spectrum is divided by its pseudo-continuum, which is computed using a 5-order polynomial fit.
The normalized spectra are used as input values into the RRNet model.

\section{A RRNet model for spectral parameter estimation}
\label{sec:Method}

To extract features from stellar spectra and restrain the negative influence from noise, we propose a neural network RRNet.
The RRNet consists of a residual module, a recurrent module and an uncertainty prediction module.
Some experiments are conducted on the reference data (section \ref{sec:Data}), RRNet shows good performance on them.
In addition, RRNet has higher accuracy and better generalization capability compared to the typical models such as \textit{StarNet} \citep{fabbro2018application, Bialek2020Assessing}.

\subsection{RRNet: Residual Recurrent Neural Network}

Earlier works \cite{bailer1997physical} and  \cite{manteiga2010anns} applied neural networks to atmospheric parameter estimation on synthetic stellar spectra. 
However, the effects of both works are limited by the data and hardware resources at that era.
\cite{fabbro2018application} proposed a convolutional neural network (\textit{StarNet}), consisting of two convolutional layers and three fully-connected layers.
The \textit{StarNet} is trained by ASSET \citep{koesterke2008center} synthetic spectra and APOGEE observed spectra, and then used in estimating stellar parameters from APOGEE observed spectra. 
Subsequently, \cite{Bialek2020Assessing} improved \textit{StarNet} using deep ensembling to give \textit{StarNet} the ability to predict parameter uncertainty.
\cite{wang2020spcanet} developed a residual-like network (\textit{SPCANet}) consisting of three convolutional layers and three fully-connected layers, and applied it to LAMOST DR7 medium-resolution spectra for stellar parameter and chemical abundance estimation. 

\begin{figure}[t]
    \centering
    \includegraphics[width=0.60\linewidth]{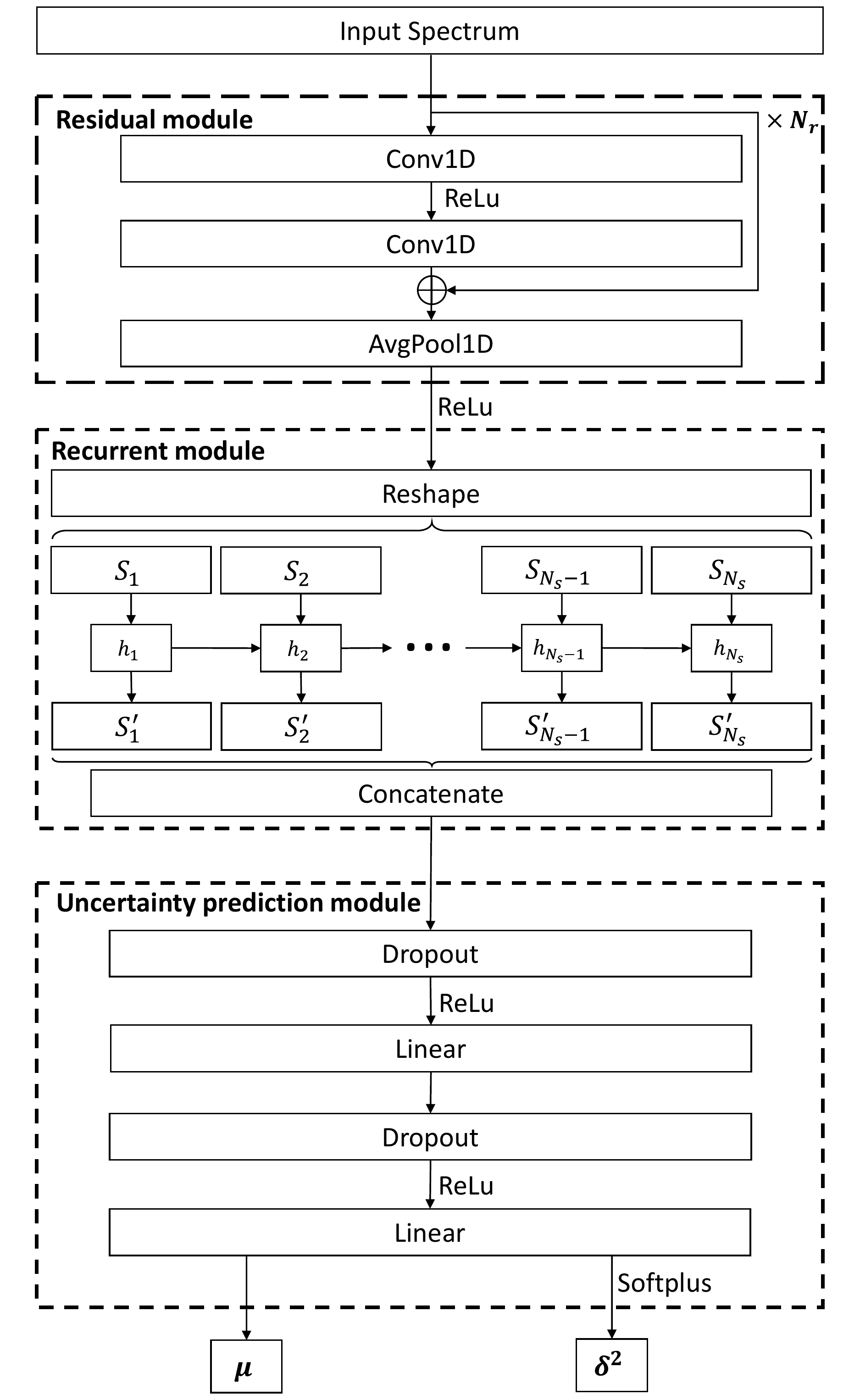}
    \caption{
    A diagram of the RRNet model.
    The input layer is a pre-processed spectrum (section \ref{sec:Data:preprocess}).
    Immediately following the input layer are $N_r$ residual blocks. 
    The residual blocks share a common structure and are used to reinforce the spectral features and restrain the negative effects from noises and irrelevant components based on their correlations with the parameter to be estimated.
    Subsequently, the spectrum is reshaped into sequence data $S$ of length $N_s$, and further processed by a recurrent layer to obtain $S'$.
    The recurrent module extracts spectral information by analyzing the correlations between the spectral features on various wavelength subbands and fusing their information.
    The final step is to establish a mapping from the spectral feature to the Probability Density Function (PDF) of stellar parameters and elemental abundances.}
    \label{fig:figure1}
\end{figure}

The above-mentioned models only employ a few layers of convolutional computations and fully-connected computations to approximate the mapping from spectra to stellar parameters.
Unfortunately, there is much space to increase the quality of the extracted features based on the experiments in section \ref{sec:Method:Model_evaluation}. 
To this end, we develop a Residual Recurrent Neural Network (RRNet) specifically for extracting spectral features and used it for estimating stellar parameters from the medium-resolution spectra in LAMOST.
The RRNet mainly consists of three kinds of modules: residual module, recurrent module and uncertainty prediction module.
The structure of RRNet is shown in Figure \ref{fig:figure1}.

\newpage
\textbf{Residual module}: 
The residual module is to explore the correlation between the spectral features and the reference labels (the parameters to be estimated).
Based on this correlation, not only the features sensitive to the parameters are enhanced, but also the negative effects from noises or irrelevant components are restrained.
The residual module contains $N_r$ residual blocks \citep{he2016deep} with the same structure.
A typical configuration in residual module is the use of batch normalization (BN) to accelerate the training procedure by reducing internal covariate shift \citep{he2016deep,Ioffe2015Batch}. 
However, our experiments on stellar parameter estimation show much more fluctuations in the training procedure on the implementation with BN than that without BN.
As a result, the training time is increased approximately  127\%, and there is some reduction on the accuracy of the spectral parameters from the learned model. 
These phenomena possibly come from the existences of a certain number of bad pixels in many of the LAMOST spectra. 
The fluxes of the bad pixels are usually enormously deviated from the normal pixels and have evident negative influences on the estimations of mean and variance in BN \citep{Ioffe2015Batch}.
Therefore, this work removed the BN from the traditional residual blocks.
In addition, our stellar parameter estimation experiments show better performance on the implementation with an Average-pooling layer than that with a Max-pooling layer.
Therefore, the proposed residual module used an average-pooling layer in each residual block.

\textbf{Recurrent module}: The recurrent module is to extract the features of various subbands from the spectrum and analyze the correlation between them.
This correlation enables maximum information recovery and noise suppression.
Since neither convolutional networks nor fully-connected networks can achieve this kind of information extraction across subbands.
For this reason, the spectral vectors are reshaped into sequence data $S=\{S_1,S_2,\cdots,S_{N_s-1}, S_{N_s}\}$ of length $N_s$,  and then $S$ is processed cross wavelength by a recurrent layer to obtain $S'$.
Through this processing, RRNet can achieve information recovery and noise suppression from the spectral features on different wavelength.

\textbf{Uncertainty prediction module}:
Following \cite{Bialek2020Assessing}, we used three fully-connected layers to predict the Probability Density Function (PDF) of stellar parameters and chemical abundances. 
The PDF is approximated using a Gaussian distribution. Therefore, RRNet just needs to output the estimation of the mean $\mu$ and variance $\sigma^2$.
In addition, we added Dropout layers \citep{Hinton2012Improving} into the Uncertainty prediction module to avoid overfitting in the model training process.
The Dropout deals with overfitting by reducing co-adaptations between neurons (especially in the same layer) and restraining the relying of a neuron on the presence of particular other neurons. 
It should be noted that to ensure the stability of the model output, Dropout is only enabled in the training phase and is disabled in the inference phase.

\subsection{Model training}
\label{sec:Method:Model_training}

The reference set is randomly divided into a training set, a validation set and a test set at the ratio of 7:1:2. 
The three data sets respectively consist of  80,812 spectra from 19,996 stars, 11,473 spectra from 2,852 stars, and  23,198 spectra from 5,705 stars.
The training set is used for the training of the RRNet model (see section \ref{sec:Method:Model_training}), the validation set is used for the selection of the model hyperparameters (see section \ref{sec:Method:Model_selection}), and the test set is used for the evaluation of the model (see section \ref{sec:Method:Model_evaluation}).
In LAMOST observations, some spectra are from the same sources. 
If the spectra from the same source simultaneously appear in the training set, validation set and test set, there is the possibility of reducing the objectivity of the model evaluation. 
Therefore, this work divides the reference set based on the sources instead of the spectra.

In order to enable the RRNet model to predict the PDF of physical parameters, consistent with \cite {Bialek2020Assessing}, we use the negative log-likelihood of the normal distribution as the loss function of the model, as follows:
\begin{equation}
    \label{equ:equation2}
    Loss(\mathbf{x}|\theta) = \frac{\log{\sigma^2_{\theta}(\mathbf{x})}}{2} + \frac{(y-\mu_{\theta}(\mathbf{x}))^2}{2\sigma^2_{\theta}(\mathbf{x})},
\end{equation}
where $\mathbf{x}, y$ are the input spectra and the corresponding reference labels, $\mu_{\theta}(\mathbf{x}), \sigma^2_{\theta}(\mathbf{x})$ are the mean and variance of the Gaussian distribution predicted by the model, and $\theta$ is the parameter of the RRNet model to be optimized. 
To learn the model parameter $\theta$, the Adam \citep{kingma2014adam} is used as the optimizer to speed up the convergence of the learning procedures.
And during the training process, a data augmentation by disturbing a reference spectrum with Gaussian noise can improve the robustness of the model.

Finally, in order to accurately estimate the PDF of the physical parameters, $M$ instances of the RRNet are trained with different random initializations ($M=6$ in this paper).
The mean of ensembling $\hat \mu(\mathbf{x})$ is obtained by averaging the predicted mean of $M$ RRNet instance models.
The variance of ensembling $\hat \sigma^2(\mathbf{x})$ is determined by the following equation:
\begin{equation}
    \label{equ:equation3}
    \hat\sigma^2(\mathbf{x}) = \frac{1}{M} \sum_{i=1}^M{(
    \sigma^2_{\theta_i}(\mathbf{x})+ \mu^2_{\theta_i}(\mathbf{x})
    ) 
    -\hat\mu^2(\mathbf{x})}.
\end{equation}

\begin{deluxetable*}{lcccc|cccc}
\tablenum{1}
\tablecaption{Experimental results for determining the  hyperparameters $N_r$ and $N_s$ of RRNet. These two hyperparameters respectively represent the numbers of residual blocks and wavelength sub-bands in RRNet. The model performance is measured using the Mean Absolute Error (MAE, defined in equation \ref{equ:equation4}) and computed from the validation set.}
\label{tab:table1}
\tablewidth{0pt}
\tablehead{
\colhead{Labels} & \multicolumn{8}{c}{MAE} \\ 
\cline{2-9} 
\colhead{} & \colhead{$N_r=1$}   &  \colhead{$N_r=2$} &  \colhead{$N_r=3$}  &  \colhead{$N_r=4$} & \colhead{$N_s=5$}   &  \colhead{$N_s=20$} &  \colhead{$N_s=40$}  &  \colhead{$N_s=60$} 
}
\decimalcolnumbers
\startdata
$T_{\rm eff}$ & 55.8792 & 52.4291         & \textbf{51.8167} & 52.4102         & 52.3478 & 52.6161         & \textbf{51.8167} & 51.7086         \\
$\log \, g$        & 0.0872  & \textbf{0.0793} & 0.0794           & 0.0812          & 0.0804  & 0.0813          & \textbf{0.0794}  & 0.0808          \\
{[}Fe/H{]}          & 0.0327  & 0.0314          & \textbf{0.0309}  & 0.0313          & 0.0315  & 0.0314          & 0.0309           & \textbf{0.0308} \\
{[}C/H{]}           & 0.0534  & 0.0522          & \textbf{0.0511}  & 0.0512          & 0.0630  & 0.0513          & \textbf{0.0511}  & 0.0514          \\
{[}N/H{]}           & 0.0763  & 0.0749          & \textbf{0.0738}  & 0.0739          & 0.0847  & 0.0739          & \textbf{0.0738}  & 0.0740          \\
{[}O/H{]}           & 0.0802  & 0.0787          & \textbf{0.0781}  & 0.0782          & 0.0888  & 0.0787          & \textbf{0.0781}  & \textbf{0.0781} \\
{[}Mg/H{]}          & 0.0372  & 0.0360          & 0.0357           & \textbf{0.0355} & 0.0421  & 0.0360          & \textbf{0.0357}  & 0.0358          \\
{[}Al/H{]}          & 0.0487  & 0.0470          & \textbf{0.0462}  & 0.0466          & 0.0539  & 0.0468          & \textbf{0.0462}  & 0.0468          \\
{[}Si/H{]}          & 0.0351  & 0.0337          & \textbf{0.0334}  & 0.0334          & 0.0406  & 0.0336          & \textbf{0.0334}  & 0.0336          \\
{[}S/H{]}           & 0.0593  & 0.0587          & \textbf{0.0580}  & 0.0588          & 0.0655  & 0.0588          & \textbf{0.0580}  & 0.0583          \\
{[}K/H{]}           & 0.0897  & 0.0902          & 0.0892           & \textbf{0.0890} & 0.0959  & \textbf{0.0891}      & 0.0892  & 0.0897          \\
{[}Ca/H{]}          & 0.0415  & 0.0401          & \textbf{0.0391}  & 0.0393          & 0.0472  & 0.0392          & \textbf{0.0391}  & 0.0395          \\
{[}Ti/H{]}          & 0.0747  & 0.0732          & 0.0732           & \textbf{0.0731} & 0.0788  & 0.0733          & \textbf{0.0732}  & 0.0729          \\
{[}Cr/H{]}          & 0.0671  & 0.0653          & \textbf{0.0651}  & 0.0652          & 0.0719  & \textbf{0.0651} & \textbf{0.0651}  & 0.0653          \\
{[}Mn/H{]}          & 0.0707  & 0.0695          & \textbf{0.0688}  & 0.0691          & 0.0763  & 0.0689          & \textbf{0.0688}  & \textbf{0.0688} \\
{[}Ni/H{]}          & 0.0425  & 0.0415          & \textbf{0.0411}  & 0.0417          & 0.0482  & 0.0412          & \textbf{0.0411}  & 0.0412          \\
{[}Cu/H{]}          & 0.1349  & 0.1339          & \textbf{0.1334}  & 0.1345          & 0.1429  & 0.1341          & \textbf{0.1334}  & 0.1336 \\
\enddata
\end{deluxetable*}

\subsection{Model selection}
\label{sec:Method:Model_selection}

In the application of machine learning, the choice of model hyperparameters have strong influences on model performance.
The depth of the RRNet is determined by the number of residual blocks.
After the residual blocks, the information of a spectrum is represented using a vector $v$ and further processed by the recurrent module in RRNet (Fig. \ref{fig:figure1}).
In the recurrent module, the $v$ is divided into a series of wavelength sub-bands (Fig. \ref{fig:figure1}). 
The numbers of residual blocks and the wavelength sub-bands are represented by $N_r$ and $N_s$ respectively. They are the hyperparameters of RRNet.
To explore the effects of the hyperparameters $N_r$ and $N_s$ on model performance, we compared the RRNet with its variants. Various variants of the RRNet are designed using different configurations of $N_r$ and $N_s$.  
In these experiments, the performance measures are computed on the validation set for determining the model hyperparameters $N_r$ and $N_s$.
This work used the Mean Absolute Error (MAE) to evaluate the performance of a stellar atmospheric parameter estimation model as follows:
\begin{equation}
    \label{equ:equation4}
    MAE = \frac{1}{n} \sum_{i=1}^n {|y_i - \hat y_i|},
\end{equation}
where $\hat{y_i}$ and $y_i$ denote the predicted value by RRNet and the corresponding reference value from a spectrum $\mathbf{x}_i$ respectively, and $n$ denotes the number of samples. 
The experimental results are presented in Table \ref{tab:table1}.

It is shown that the model performance is improved with the increases of $N_r$ and $N_s$.
However, after these hyperparameters reach certain thresholds, the performance improvement will be trivial or even degrade.
The above-mentioned phenomena indicate that more residual blocks and more wavelength sub-bands in an RRNet model help to improve the spectral information extraction capability and increase the parameter estimation performance to a certain degree.
More residual blocks result in an RRNet with more complexity and more model parameters to be optimized.
This kind of models need a training data set with more reference spectra.
However, the scale of the training set is difficult to be expanded without any limitations in a real application.
Therefore, the performance of the RRNet firstly is improved gradually with the increase of $N_r$, and then decreases in case that the complexity of the RRNet exceeds that supporting from the available training data.
Therefore, the performance of the RRNet can be improved further by adding more residual blocks if a training set with a larger scale can be obtained.

Another characteristic of the RRNet is that the model can exploit spectral features and restrain the negative influences from noise based on CBE between the observations on different wavelengths (Fig. \ref{fig:figure1}). 
Two non-directly adjacent sub-bands can indirectly communicate with each other through a series of directly adjacent sub-bands between them, and conduct belief enhancement.
In the recurrent module, a series of sub-bands can be numbered and indexed by integers from 1 to $N_s$  from left to right, and the distance between each two sub-bands can be measured by the difference between their indexes (Fig. \ref{fig:figure1}).
The further the distance between two non-direct adjacent sub-bands, the weaker the communication and information fusion ability between them.
In the case of the number of sub-bands changing from 1, to 2, to 3, similarly to more, the amount of sub-band pairs with a directly adjacent relationship or small distance from each other increases gradually. 
Therefore, the CBE ability and prediction effect of RRNet are improved gradually in this procedure.
At the same time, with the increase of $N_s$, more and more long-distance sub-band pairs appear.
This kind of sub-band pairs results in poor communication between them and unsatisfactory belief enhancement. 
Therefore, the performance of the RRNet increases firstly and then decreases in the case of $N_s$ gradually increasing. 
Ultimately, this work utilized the configuration $N_r=3$ and $N_s=40$ based on the experimental investigations.



\begin{figure}
  \begin{minipage}[ht]{1\linewidth}
    \centering
    \includegraphics[width=1\linewidth]{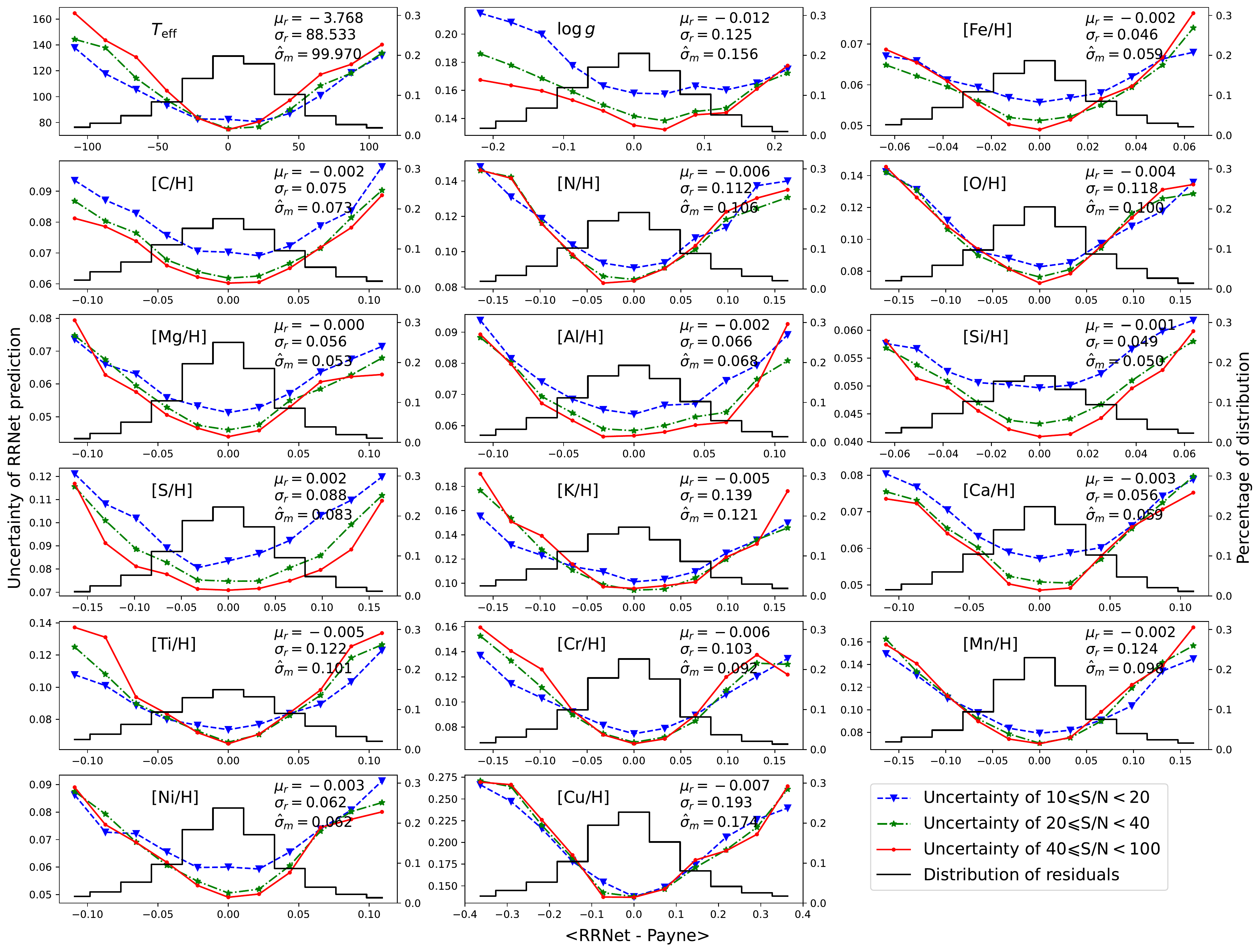}
    \caption{
    The consistencies between RRNet prediction and APOGEE-\textit{Payne} catalog and the robustness against data quality.
    The black line is the distribution curve of the inconsistencies (difference) between RRNet predictions and APOGEE-\textit{Payne} catalog, and three colored curves (blue, green, and red) present the uncertainties (standard deviation) of the inconsistencies.
    $\mu_r$ and $\sigma_r$ are respectively the mean and standard deviation of the difference, $\hat \sigma_m$ is the mean of the $1\sigma$ uncertainty predicted by RRNet.}
    \label{fig:figure2}
  \end{minipage}

  \begin{minipage}[ht]{1\linewidth}
    \centering
    \includegraphics[width=1\linewidth]{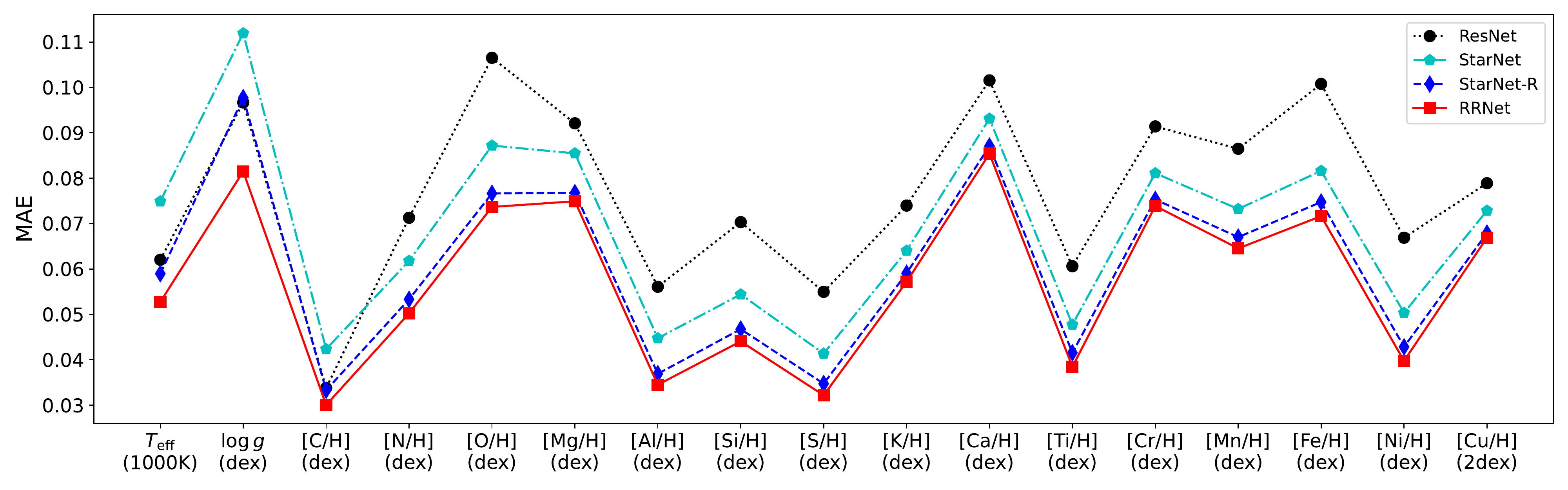}
    \caption{Performance comparisions: RRNet, \textit{ResNet} , \textit{StarNet}, and \textit{StarNet-R}. The performance is evaluated based on the MAE (equation \ref{equ:equation4}). The \textit{StarNet-R} is a variant of \textit{StarNet} with recurrent module same with the RRNet.}
    \label{fig:figure3}
  \end{minipage}
\end{figure}

\begin{figure}[t]
    \centering
    \includegraphics[width=1\linewidth]{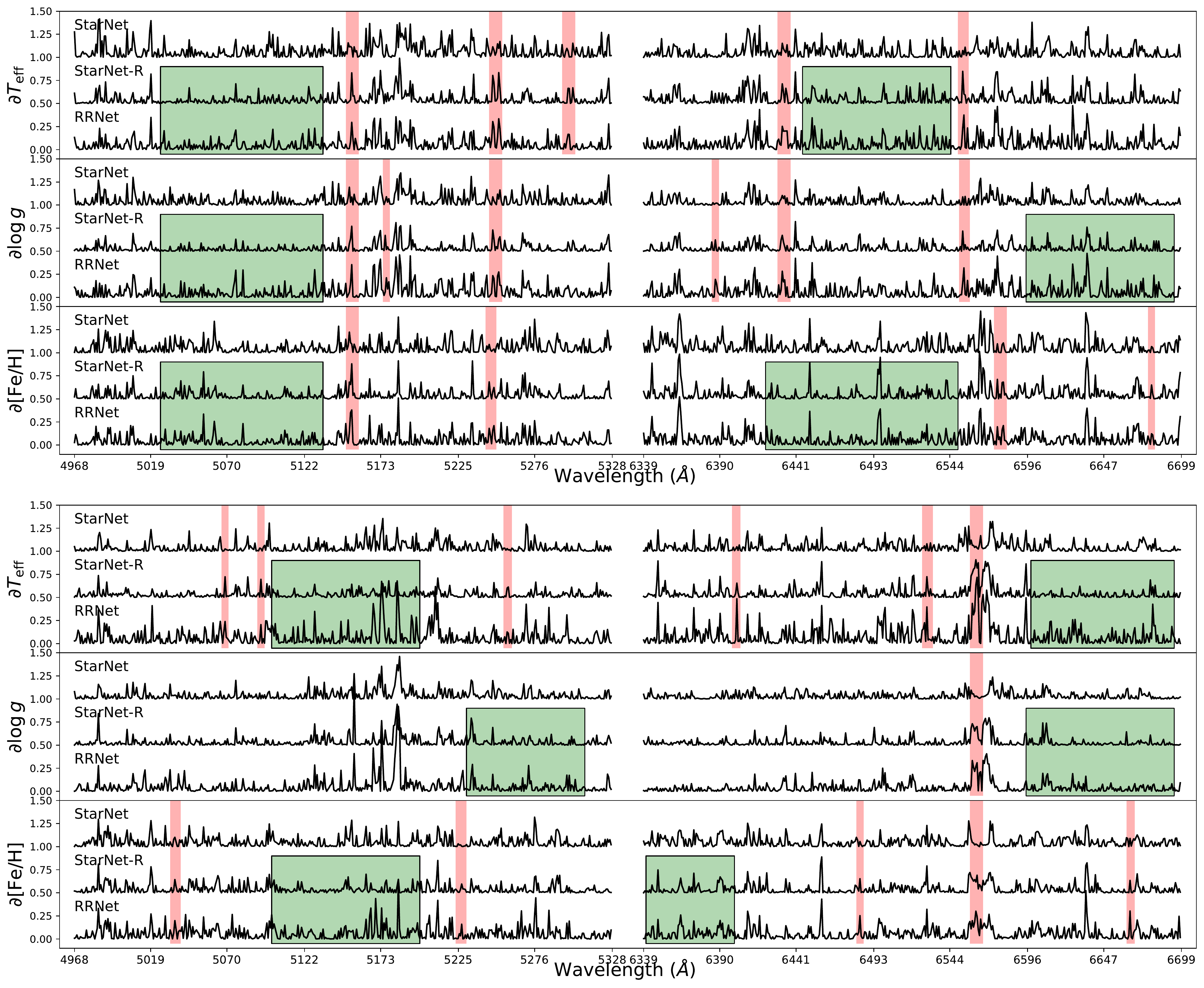}
    \caption{
    The average partial derivatives of the stellar parameters by \textit{StarNet}, \textit{StarNet-R} and RRNet models over the cool  and hot stars. 
    The upper and lower panels present the average partial derivatives respectively for cool stars ($T_{\rm eff} \in \rm [4000, 4300]$ K) and hot stars ($T_{\rm eff} \in \rm [6000, 6300]$ K).
    Each average partial derivative is computed from 1000 randomly selected spectra in the test set.
    The sub-bands shaded in red indicate that the model can extract more weak spectral features with the recurrent module, and the sub-bands with borders shaded in green show that the model can perform more detailed features with the residual module.}
    \label{fig:figure4}
\end{figure}

\subsection{Model evaluation}
\label{sec:Method:Model_evaluation}

After determining the optimal hyperparameters of the model, we demonstrate the performance of RRNet on the test set and compare it with other models.
Figure \ref{fig:figure2} shows the distribution of the inconsistencies, and the dependence of the $1\sigma$ uncertainty predicted by RRNet on S/N and inconsistencies.
The inconsistency refers to the difference between the RRNet predictions and the APOGEE-\textit{Payne} catalog on the test set.
On the whole, the RRNet predictions have high accuracy and precision.
The precision of the stellar parameters ($T_{\rm eff}$ and $\log \, g$) are 88 K and 0.13 dex respectively, elements C, Mg, Al, Si, Ca, Fe, Ni are 0.05 dex to 0.08 dex, and N, O, S, K, Ti, Cr, Mn are 0.09dex to 0.14 dex, while that of Cu is 0.19 dex.
In addition, although the quality of a spectrum helps improve the parameter estimation performance (small bias and standard deviation) on the whole, the standard deviation of the RRNet prediction varies in a small range.
Therefore, the RRNet are robust against noises and disturbances.
The robustness benefits from the CBE of the recurrent module, which enables the RRNet model to fuse the feature information from different wavelength sub-bands and restrain the negative influences from noise and disturbances based on the correlation between different sub-bands.

To further demonstrate the performance of RRNet, we compared RRNet with models such as \textit{StarNet}.
\cite{fabbro2018application} mentions that \textit{StarNet} can be tried to apply to other spectral studies.
Therefore, this work investigated its application on LAMOST DR7 by training the \textit{StarNet} using the training set same with RRNet, and compared its results with our model.
In addition, to verify the effectiveness of the combination of residual module and recurrent module, we constructed a model \textit{StarNet-R} by adding the recurrent module same with RRNet into the \textit{StarNet}, and compared the RRNet with \textit{ResNet} and \textit{StarNet-R}.
Figure \ref{fig:figure3} shows the MAE (equation \ref{equ:equation4}) of parameter prediction for RRNet, \textit{ResNet}, \textit{StarNet} and \textit{StarNet-R} on the test set.

It is shown that RRNet has significant advantages over the other three models (Figure \ref{fig:figure3}).
The performance of \textit{ResNet} is lower than that of RRNet because \textit{ResNet} only extracts flux features longitudinally from the spectrum based on the correlation between the fluxes and the parameter to be estimated,  and lacks the capability of CBE.
The \textit{StarNet} consists of several convolutional layers and fully-connected layers.
It is shown that this method does not work as well as RRNet on LAMOST spectra.
After adding a recurrent module, a variant of \textit{StarNet} is obtained and referred to as \textit{StarNet-R}.
Therefore, the \textit{StarNet-R} has the capability of CBE and shows a performance superior to the original \textit{StarNet} (Figure \ref{fig:figure3}). 
In RRNet, there are both the residual module and recurrent module. Therefore, the RRNet has better performance on 1) feature extraction for the residual module than \textit{StarNet} and \textit{StarNet-R}; 2) information recovery and noise \& disturbances abatement for CBE from recurrent module than \textit{StarNet} and \textit{ResNet}.
Therefore, it is shown that the RRNet is more accurate and stable compared with other models.

To visualize why the RRNet modules are effective, Figure \ref{fig:figure4} shows the average partial derivatives of the stellar parameters by the \textit{StarNet}, \textit{StarNet-R}, and RRNet.
The upper and lower panels of the figure represent the spectral average partial derivatives from 1000 randomly selected cool stars ($T_{\rm eff} \in \rm [4000, 4300]$ K) and 1000 hot stars ($T_{\rm eff} \in \rm [6000, 6300]$ K) in the test set, respectively.
The results show that, on the whole, three models are consistent with each other very well on their average partial derivatives.

The comparisons on (RRNet, \textit{StarNet}, \textit{StarNet-R}) and their visualizations are further summarized. 
1) Both \textit{StarNet-R} and RRNet have a recurrent module which enhances some hidden sub-band features compared to \textit{StarNet} without the recurrent module (See Figure \ref{fig:figure4} for the sub-bands shaded in red).
These phenomena indicate that the recurrent module does help recover the spectral features through CBE and thus improve the performance of the model.
2) Both \textit{StarNet-R} and RRNet have recurrent modules, the difference is that \textit{StarNet-R} uses a traditional convolutional neural network to extract spectral features while RRNet uses the superposition of multiple residual blocks to extract features.
The results show that the residual module reinforces more dense features than the traditional convolutional neural network (See Figure \ref{fig:figure4} for the sub-bands with borders shaded in green).
These phenomena indicate that the residual module does provide stronger feature extraction performance than the traditional convolutional neural network.

\begin{figure}[t]
    \centering
    \includegraphics[width=1\linewidth]{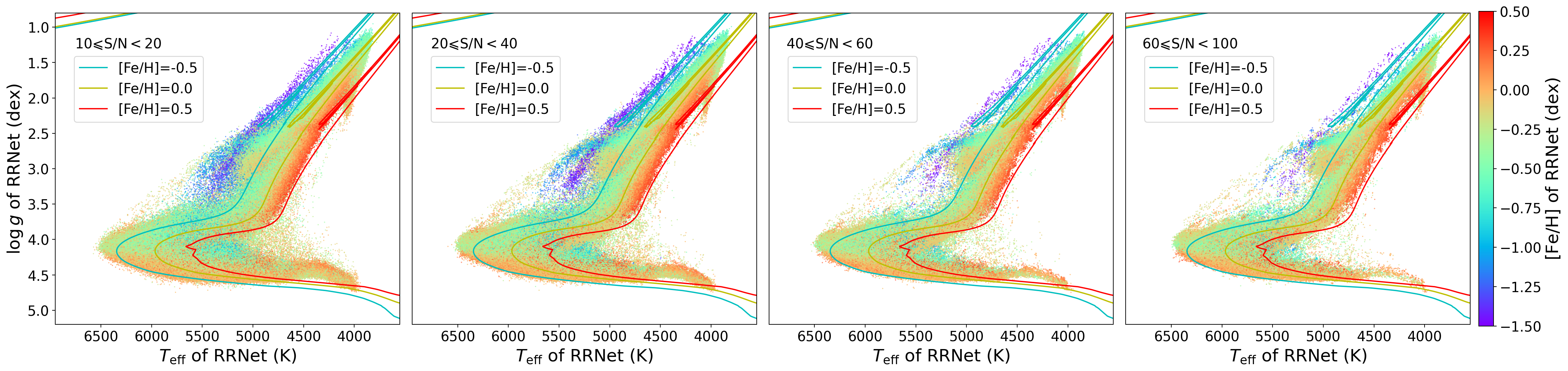}
    \caption{Distribution of stellar parameters from LAMOST-RRNet catalog. The color characterizes [Fe/H], and the three isochrones represent the MIST stellar evolution tracks with stellar ages of 7 Gyr ($\rm [Fe/H]$ of -0.5, 0.0, and 0.5 respectively). The S/N intervals are $\rm 10 \leq S/N < 20$, $\rm 20 \leq S/N < 40$, $\rm 40 \leq S/N < 60$, and $\rm 60 \leq S/N < 100$, respectively. }
    \label{fig:figure5}
\end{figure}

\begin{figure}[t]
    \centering
    \includegraphics[width=1\linewidth]{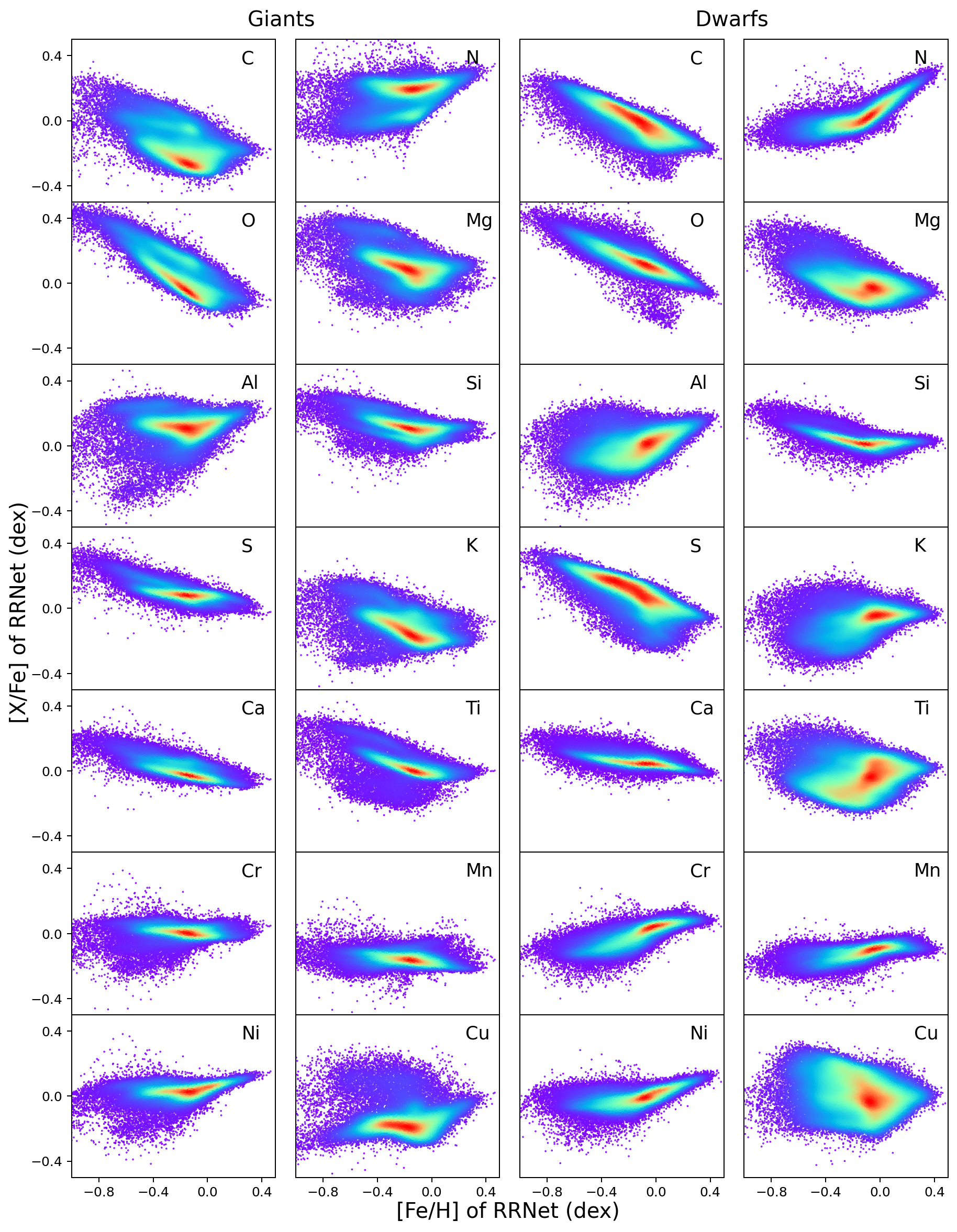}
    \caption{Distribution of elemental abundances [X/Fe] relative to [Fe/H] from LAMOST-RRNet catalog. The left panel is for giants ($\log \, g < 4$) and the right panel is for dwarfs ($\log \, g > 4$).}
    \label{fig:figure6}
\end{figure}

\section{Application on LAMOST DR7}
\label{sec:Results}

In this section, we applied RRNet to the medium-resolution spectra from LAMOST DR7, computed a LAMOST-RRNet catalog  and evaluated the catalog.
The establishment of the LAMOST-RRNet catalog and the related validation are further described in the following subsections.

\subsection{LAMOST DR7 parameter estimation}

After training and testing the RRNet model (sections \ref{sec:Method:Model_training} and \ref{sec:Method:Model_evaluation}), the stellar parameters and elemental abundances are derived for medium-resolution spectra from LAMOST DR7 by RRNet.
Based on the distribution range of the stellar parameters in the reference set, we kept only the LAMOST DR7 spectra with LASP \citep{luo2015first, Wu2011Automatic} estimation $T_{\rm eff} \in \rm [3500, 7000]$ K, and processed these spectra using the pre-processing procedures in section \ref{sec:Data:preprocess}.
Ultimately, the LAMOST-RRNet catalog is obtained by RRNet.
This catalog contains stellar atmospheric parameters, chemical abundances, and corresponding $1\sigma$ uncertainties for 2,377,510 medium-resolution spectra in LAMOST DR7 estimated by RRNet.

The $T_{\rm eff}-\log \, g$ distributions for the spectra with S/N in multiple intervals are shown in Fig. \ref{fig:figure5}, and three MIST stellar isochrones \citep{dotter2016mesa, choi2016mesa} with stellar ages of 7 Gyr are added to the figure for reference.
The stellar parameters estimated by RRNet are consistent with the three MIST stellar isochrones, and the distribution of $T_{\rm eff}-\log \, g$ becomes ``cleaner" as $\rm S/N$ increases.
The density distributions of [X/Fe] relative to [Fe/H] for giant and dwarf stars are shown in Figure \ref{fig:figure6}.
In general, the elemental abundances estimated by RRNet are more dense, especially for Si, S, Ca, and Ni.
In addition, the $\alpha$ elements (O, Mg, Si, S, Ca, Ti) show a more obvious bimodal structure on the giant star samples.

\begin{deluxetable*}{lcccccccccccc}
\tablenum{2}
\tablecaption{Some comparisons of the LAMOST-RRNet catalog with other catalogs.}
\label{tab:table2}
\tablewidth{0pt}
\tablehead{
\colhead{Labels} & \multicolumn{3}{c}{SPCANet-Payne} & \multicolumn{3}{c}{RRNet-Payne}  & \multicolumn{2}{c}{RRNet-GALAH}  & \multicolumn{2}{c}{RRNet-ASPCAP}   & \multicolumn{2}{c}{Payne-ASPCAP} \\ 
\cline{2-13} 
\colhead{} & \colhead{$\mu_r$}   & \colhead{$\sigma_r$}    & \colhead{MAE}  & \colhead{$\mu_r$}   & \colhead{$\sigma_r$}    & \colhead{MAE}  & \colhead{$\mu_r$}   & \colhead{$\sigma_r$}    & \colhead{$\mu_r$}   & \colhead{$\sigma_r$} & \colhead{$\mu_r$}   & \colhead{$\sigma_r$}
}
\decimalcolnumbers
\startdata
$T_{\rm eff}$       & 17.73                 & 87.08                 & 63.74                 & -0.63              & 73.21                 & 43.07 & -52.87               & 131.44               & -58.09 & 108.31 & -84.49 & 99.65 \\
$\log \, g$        & 0.000                 & 0.126                 & 0.092                 & -0.008             & 0.106                 & 0.068 & 0.063                & 0.156                & 0.009  & 0.161  & 0.023  & 0.129 \\
{[}Fe/H{]}         & 0.003                 & 0.045                 & 0.033                 & -0.001             & 0.038                 & 0.024 & -0.027               & 0.095                & -0.031 & 0.063  & -0.027 & 0.053 \\
{[}C/H{]}          & -0.001                & 0.068                 & 0.048                 & -0.002             & 0.066                 & 0.044 & -0.123               & 0.139                & -0.159 & 0.112  & -0.186 & 0.121 \\
{[}N/H{]}           & 0.003                 & 0.104                 & 0.073                 & -0.001             & 0.101                 & 0.066 & $\cdots$ & $\cdots$ & -0.029 & 0.160  & -0.012 & 0.144 \\
{[}O/H{]}          & 0.003                 & 0.098                 & 0.067                 & -0.003             & 0.110                 & 0.069 & -0.070               & 0.179                & -0.038 & 0.101  & -0.048 & 0.121 \\
{[}Mg/H{]}         & -0.006                & 0.050                 & 0.037                 & -0.001             & 0.044                 & 0.028 & -0.036               & 0.137                & 0.002  & 0.098  & 0.027  & 0.091 \\
{[}Al/H{]}         & -0.004                & 0.062                 & 0.045                 & -0.001             & 0.056                 & 0.037 & -0.074               & 0.113                & 0.006  & 0.135  & 0.032  & 0.131 \\
{[}Si/H{]}        & -0.006                & 0.047                 & 0.034                 & -0.001             & 0.041                 & 0.026 & -0.020               & 0.096                & 0.027  & 0.084  & 0.054  & 0.081 \\
{[}S/H{]}          & -0.001                & 0.082                 & 0.057                 & 0.002              & 0.082                 & 0.054 & $\cdots$ & $\cdots$ & -0.016 & 0.109  & -0.017 & 0.137 \\
{[}K/H{]}          & $\cdots$                & $\cdots$                & $\cdots$                & -0.003             & 0.130                 & 0.080 & -0.193               & 0.289                & -0.163 & 0.131  & -0.166 & 0.170 \\
{[}Ca/H{]}         & -0.004                & 0.054                 & 0.039                 & -0.003             & 0.049                 & 0.033 & -0.071               & 0.125                & -0.035 & 0.078  & -0.046 & 0.078 \\
{[}Ti/H{]}         & -0.003                & 0.105                 & 0.067                 & -0.003             & 0.115                 & 0.067 & -0.091               & 0.169                & -0.008 & 0.147  & 0.001  & 0.153 \\
{[}Cr/H{]}         & 0.016                 & 0.098                 & 0.063                 & -0.002             & 0.097                 & 0.058 & 0.001                & 0.127                & 0.016  & 0.158  & 0.023  & 0.145 \\
{[}Mn/H{]}         & $\cdots$                & $\cdots$                & $\cdots$                & -0.002             & 0.105                 & 0.063 & -0.150               & 0.133                & -0.152 & 0.110  & -0.159 & 0.154 \\
{[}Ni/H{]}          & 0.001                 & 0.058                 & 0.039                 & -0.001             & 0.051                 & 0.032 & -0.016               & 0.132                & -0.030 & 0.079  & -0.023 & 0.076 \\
{[}Cu/H{]}         & 0.012                 & 0.171                 & 0.121                 & -0.002             & 0.182                 & 0.124 & -0.130               & 0.155                & -0.151 & 0.240  & -0.198 & 0.255 \\
\enddata
\tablecomments{
The $\mu_r, \sigma_r$ are the mean value and the standard deviation of the difference between the two catalogs, respectively.
}
\end{deluxetable*}

\subsection{Some comparisons with other surveys}

To verify the accuracy and precision of LAMOST-RRNet catalog, we investigated the consistency of LAMOST-RRNet catalog with the \textit{SPCANet} catalog, the APOGEE-ASPCAP DR16 catalog, and the GALAH DR3 catalog on the spectra from the common stars between them.

\cite{wang2020spcanet} proposed a residual-like neural network model (\textit{SPCANet}) to estimate stellar parameters and elemental abundances for 1,472,211 medium-resolution spectra from LAMOST DR7. 
Both \textit{SPCANet} catalog \citep{wang2020spcanet} and LAMOST-RRNet catalog are computed from parameter estimation models learned from the the spectra of the common stars between APOGEE-\textit{Payne} catalog and LAMOST observations. 
Therefore, the comparability between these two catalogs are very well.
The comparing tests are performed on the LAMOST observations from 24,347 common stars.
It should be noted that there are three kinds of estimations from the spectra of these common stars: the estimation of the \textit{SPCANet} model, the estimation of the RRNet model, and the reference values of the APOGEE-\textit{Payne} catalog.

APOGEE \citep{majewski2017apache} is a medium-high resolution (R $\thicksim$ 22500) spectroscopic survey, which uses the Sloan telescope at Apache Point Observatory in New Mexico City, USA, to achieve observations of stellar spectra.
The APOGEE spectral band covers the near-infrared 1.51-1.70 µm, and the stellar parameters and elemental abundances for most of their stellar spectra are mainly calculated by the ASPCAP \citep{perez2016aspcap}.
We cross-matched LAMOST-RRNet with the APOGEE-ASPCAP DR16 catalog, and excluded stars with STARFLAG, ASPCAPFLAG, and PARAMFLAG flags.
Finally, 127,574 LAMOST DR7 spectra from 31,356 common stars are obtained.

GALAH \citep{de2015galah} is a large-scale high-resolution (R $\thicksim$ 28000) spectroscopic survey, which uses the Anglo-Australian Telescope and the HERMES spectrograph at the Australian Observatory to observe stellar spectra.
The GALAH spectral coverage are [4713, 4903] \text{\AA}, [5648, 5873] \text{\AA}, [6478, 6737] \text{\AA} and [7585, 7887] \text{\AA}.
GALAH DR3 \citep{buder2021galah+} presented stellar parameters and elemental abundances for a total of 588,571 stars, including 383,088 dwarfs, 200,927 giants, and 4,556 unclassified stars.
We cross-matched LAMOST-RRNet with the Galah DR3 catalog, and removed the observations with $\rm flag\_sp\neq 0$ and $\rm snr\_c3\_iraf \leq 30$ as suggested by \cite{buder2021galah+}.
Finally, 74,090 LAMOST DR7 spectra from 16,582 common stars are obtained.

Compared with the \textit{SPCANet} catalog (see Table \ref{tab:table2} (SPCANet-Payne, RRNet-Payne)), LAMOST-RRNet adds the estimations for [K/H] and [Mn/H], and reduces the overall MAE by 14\% on their common parameters.
In addition, LAMOST-RRNet has smaller bias and standard deviation compared to the \textit{SPCANet} catalog. 
Therefore, RRNet has better estimation performance.
For elements Ti and Cu, the precision improvement of the RRNet model are not significant, which may be caused by the lack of stronger metal lines in the blue part from the LAMOST spectra.
In addition, the \textit{SPCANet} input spectrum has 8000 data points (fluxes), while the RRNet input spectrum has only 7200 data points (fluxes).
The less inputs result in less feature space that can be extracted by RRNet than \textit{SPCANet}. 
Compared to the GALAH and APOGEE-ASPCAP catalogs (see Table \ref{tab:table2} (RRNet-GALAH, RRNet-ASPCAP)), LAMOST-RRNet shows good consistencies with APOGEE-ASPCAP and GALAH with smaller bias.
Similar with the findings by \cite{wang2020spcanet}, for $T_{\rm eff}$, [C/H], [K/H], [Mn/H] and [Cu/H], the LAMOST-RRNet have different levels of underestimation compared to APOGEE-ASPCAP and GALAH.
The same phenomenon is shown between APOGEE-\textit{Payne} and APOGEE-ASPCAP (see Table \ref{tab:table2} Payne-ASPCAP).
This indicates that the inconsistency between RRNet and APOGEE-ASPCAP, in addition to the inconsistency in RRNet model estimation, may partly originate from the APOGEE-\textit{Payne} label.

\begin{figure}[t]
    \centering
    \includegraphics[width=1\linewidth]{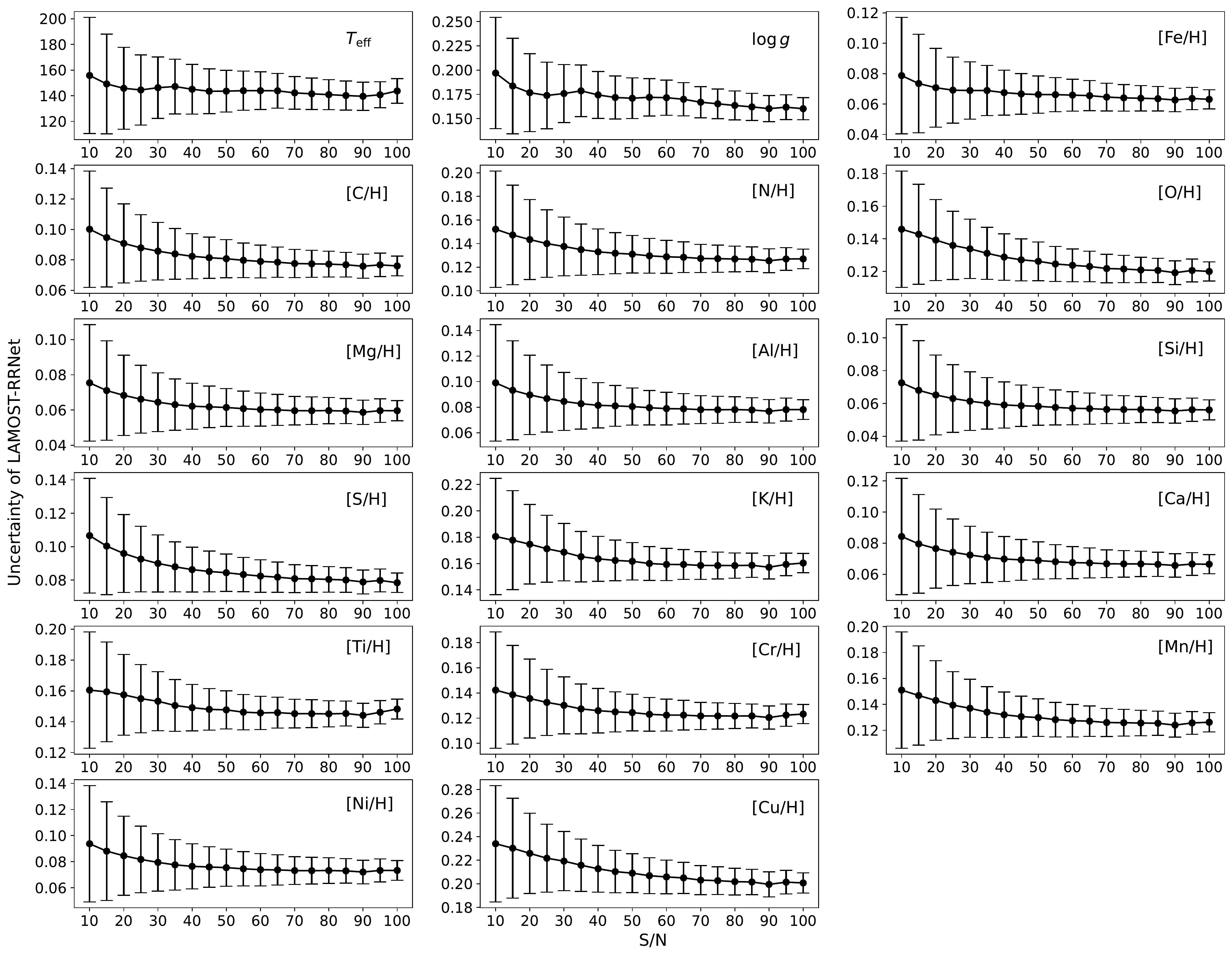}
    \caption{Dependencies of parameter estimation uncertainties on S/N. The dots indicate the uncertainty predicted by RRNet, and the length of the line segments centered on the dots indicate the uncertainty estimated from repeated observations ($>5$ times).}
    \label{fig:figure7}
\end{figure}

\subsection{Uncertainty analysis}
\label{sec:Results:Uncertainty_analysis}

The RRNet is able to predict the PDF of parameters more accurately through deep ensembling.
The uncertainty of the predicted parameters $\sigma_{pred}$ is obtained by the PDF of the parameter estimations.
In addition, in the LAMOST sky survey, some stars are observed for multiple times at various time and under different observation conditions.
Therefore, we can use this phenomenon to analyze the uncertainty caused by observation errors, this uncertainty is denoted by $\sigma_{obs}$.
Suppose we have $n_s$ repeated observations $\{\mathbf{x}_1, \cdots, \mathbf{x}_{n_s} \}$ from a source. From these observations, the RRNet gives $n_s$ estimations $\{y_1, \cdots, y_{n_s} \}$ for any stellar parameter $X$. 
The standard deviation of $\{y_1, \cdots, y_{n_s} \}$ is the corresponding uncertainty $\sigma_{obs}$.

Figure \ref{fig:figure7} shows the dependencies of the uncertainty of LAMOST-RRNet catalog on S/N.
The dots indicate uncertainty $\sigma_{pred}$ predicted by RRNet and the length of the line segment centered on the dots indicates the uncertainty $\sigma_{obs}$.
The lower uncertainty indicates the strong robustness and generalization ability of the RRNet model.
When S/N $\geq 20$, the $\sigma_{pred}$ of the parameters $T_{\rm eff}$, $\log \, g$, [Fe/H], [Cu/H] are 146 K, 0.18 dex, 0.07 dex and 0.22 dex, respectively, and those of the remaining elements are 0.06 dex $\thicksim$ 0.17 dex.
In addition, both $\sigma_{pred}$ and $\sigma_{obs}$ decrease with the increase of S/N.
This phenomenon indicates that the $\sigma_{pred}$ and $\sigma_{obs}$ are rational indicators of uncertainty.
At the same time, the changes of $\sigma_{pred}$ trend curve is mild.
Therefore, the RRNet model is stable for estimating parameters from LAMOST spectra.

\begin{figure}[t]
    \centering
    \includegraphics[width=1\linewidth]{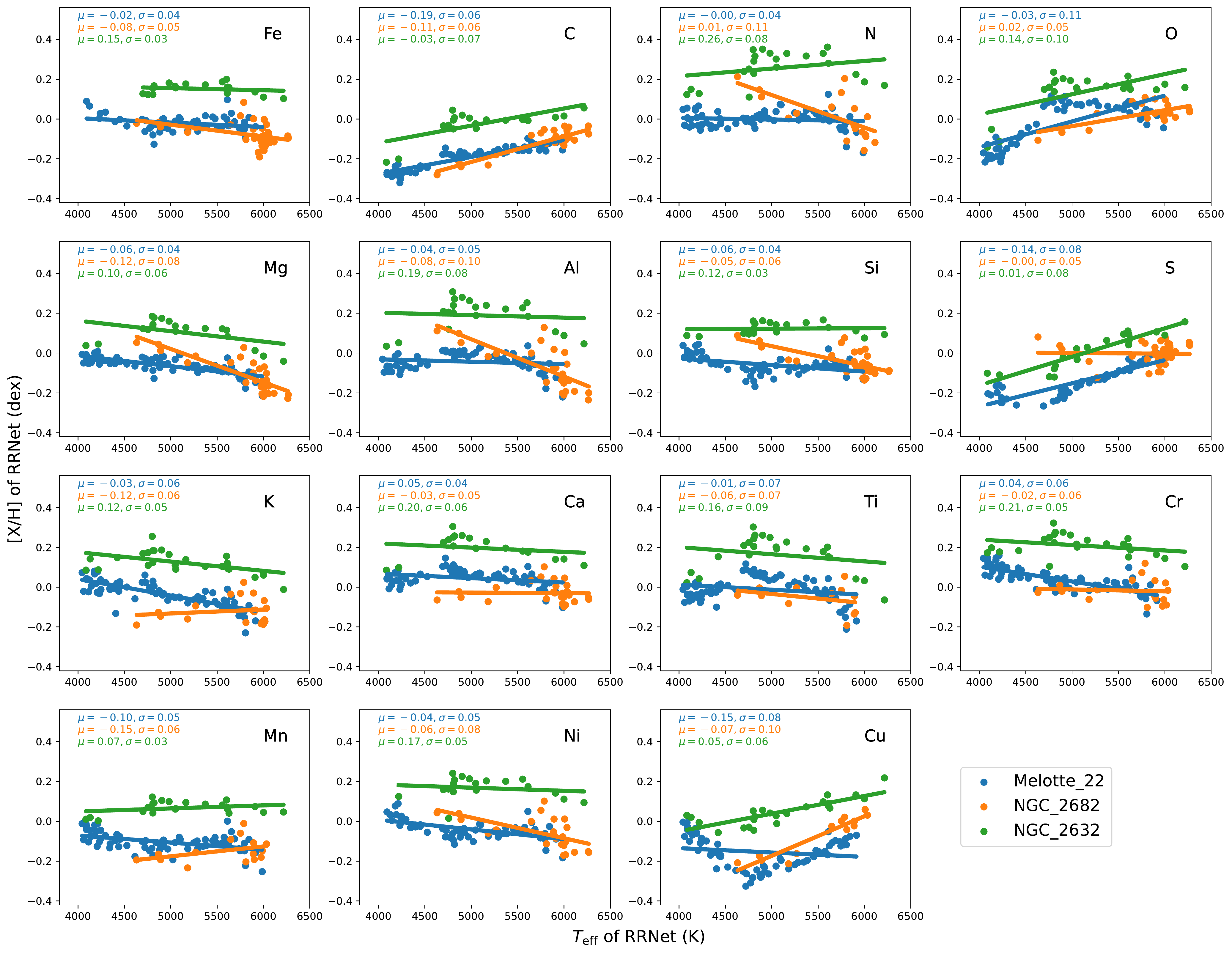}
    \caption{
    The elemental abundances (from LAMOST-RRNet) variations with $T_{\rm eff}$ in the three open clusters of Melotte 22, NGC 2682, NGC 2632. The three colors in the figure correspond to the three open clusters, and the mean {$\mu$} and standard deviation {$\sigma$} of the chemical abundances are added to each panel. The reference line is obtained by linear regression fit.}
    \label{fig:figure8}
\end{figure}

\begin{figure}[t]
    \centering
    \includegraphics[width=1\linewidth]{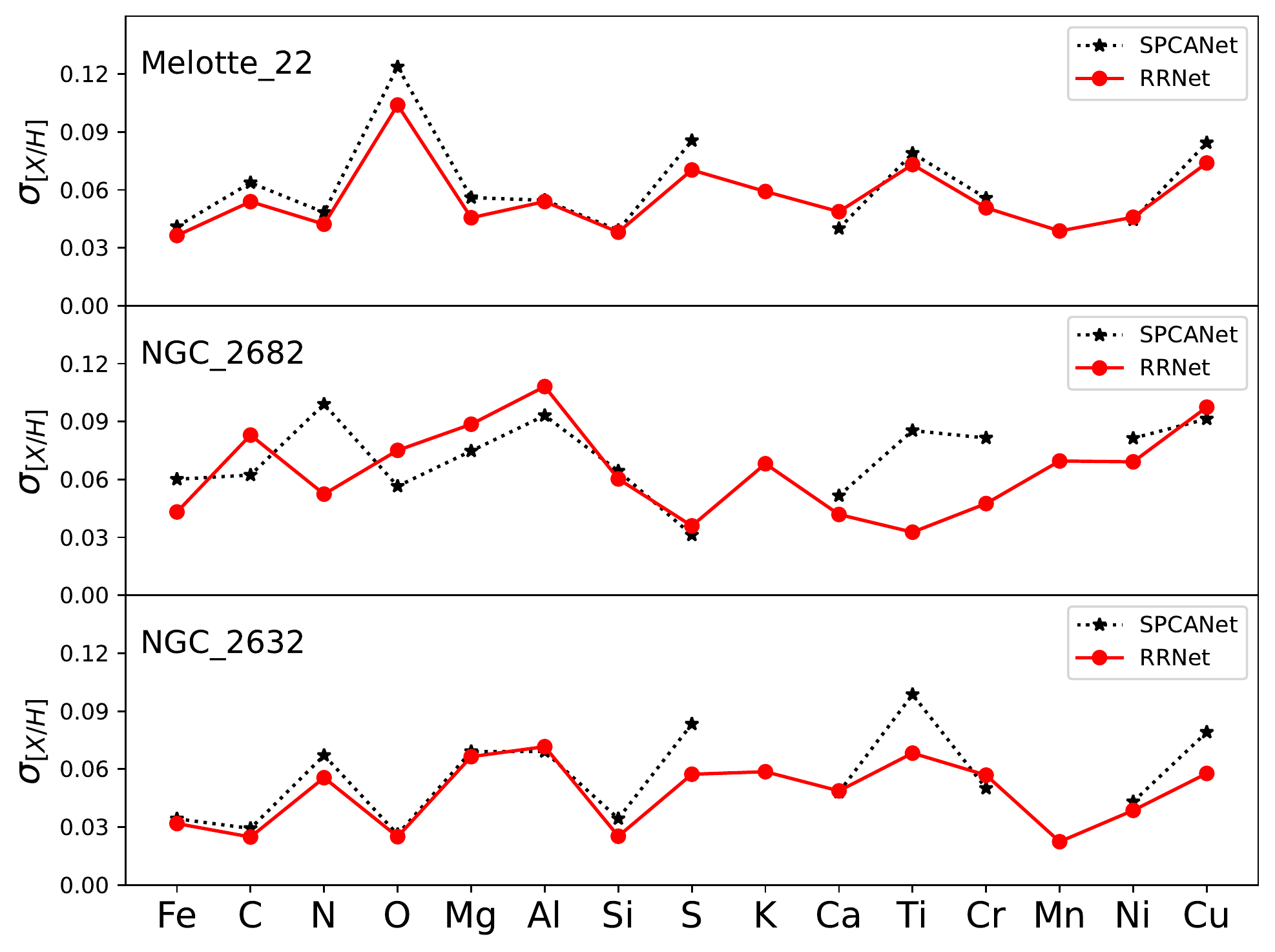}
    \caption{Some comparisons of the chemical abundances from LAMOST-RRNet and \textit{SPCANet} catalogs in the three open clusters (Melotte 22, NGC 2682, and NGC 2632).}
    \label{fig:figure9}
\end{figure}

\subsection{Test on open clusters}

To further examine the accuracy of element abundance from LAMOST-RRNet, we performed more tests on open clusters.
Since the stars in open clusters are produced almost simultaneously from gas clouds, open clusters are a stellar population of chemically homogeneous \citep{Bovy2016The, Ness2018Galactic}.
Therefore, we performed some tests using 8,811 cluster member stars published by \cite{Zhong2020Exploring}.
We selected the three open clusters (Melotte 22, NGC 2682, and NGC 2632) with the largest number of matches to LAMOST-RRNet and removed parameter estimations with large uncertainties $\sigma_{pred}$ (section \ref{sec:Results:Uncertainty_analysis}).

Figure \ref{fig:figure7} shows the dependencies of the elemental abundances (from LAMOST-RRNet) on $T_{\rm eff}$ in the above-mentioned clusters.
In agreement with \cite{ting2019payne}, the LAMOST-RRNet do not show any evident [X/H]-$T_{\rm eff}$ trend in all three aforementioned clusters, and the chemical abundances exhibit a low standard deviation.
In addition, Figure \ref{fig:figure9} shows the comparison of the chemical abundances of LAMOST-RRNet and \textit{SPCANet} catalogs in the above-mentioned clusters.
Overall, the standard deviation of chemical abundances from LAMOST-RRNet is slightly lower than that of the \textit{SPCANet} catalog.
The overall chemical homogeneity from LAMOST-RRNet on the three clusters are $0.055\pm0.017$ dex, $0.064\pm0.022$ dex and $0.047\pm0.016$ dex, respectively.
These phenomena indicate that RRNet has higher accuracy compared to \textit{SPCANet}.

\begin{deluxetable*}{lll}
\tablenum{3}
\tablecaption{The description of LAMOST-RRNet catalog. \label{tab:table3}}
\tablewidth{0pt}
\tablehead{
\colhead{Col.} & \colhead{Marker} & \colhead{Description} 
}
\decimalcolnumbers
\startdata
1 & obsid & Identification of LAMOST spectrum \\
2 & filename & Fits file name of LAMOST spectrum \\
3 & ra & R.A. of J2000 (°) \\
4 & dec & Decl. of J2000 (°) \\
5 & extname\_blue & Spectrum extension name of blue part \\
6 & extname\_red & Spectrum extension name of red part\\
7 & snr\_blue & Signal-to-noise ratio of blue part\\
8 & snr\_red & Signal-to-noise ratio of red part\\
9-10 & Teff[K], Teff[K]\_err & Effective temperature and its uncertainty (K) \\
11-12 & Logg, Logg\_err & Surface gravity and its uncertainty (dex) \\
13-14 & CH, CH\_err & [C/H] and its uncertainty (dex) \\
15-16 & NH, NH\_err & [N/H] and its uncertainty (dex) \\
17-18 & OH, OH\_err & [O/H] and its uncertainty (dex) \\
19-20 & MgH, MgH\_err & [Mg/H] and its uncertainty (dex) \\
21-22 & AlH, AlH\_err & [Al/H] and its uncertainty (dex) \\
23-24 & SiH, SiH\_err & [Si/H] and its uncertainty (dex) \\
25-26 & SH, SH\_err & [S/H] and its uncertainty (dex) \\
27-28 & KH, KH\_err & [K/H] and its uncertainty (dex) \\
29-30 & CaH, CaH\_err & [Ca/H] and its uncertainty (dex) \\
31-32 & TiH, TiH\_err & [Ti/H] and its uncertainty (dex) \\
33-34 & CrH, CrH\_err & [Cr/H] and its uncertainty (dex) \\
35-36 & MnH, MnH\_err & [Mn/H] and its uncertainty (dex) \\
37-38 & FeH, FeH\_err & [Fe/H] and its uncertainty (dex) \\
39-40 & NiH, NiH\_err & [Ni/H] and its uncertainty (dex) \\
41-42 & CuH, CuH\_err & [Cu/H] and its uncertainty (dex) \\
43 & flag & Recommended flag, 1 for good\\
\enddata
\tablecomments{
The complete catalog is available for download at \\
\href{https://github.com/Chan-0312/RRNet/releases}{https://github.com/Chan-0312/RRNet/releases}.}
\end{deluxetable*}

\subsection{LAMOST DR7 RRNet catalog}

Finally, the LAMOST-RRNet catalog of stellar atmospheric parameters and elemental abundances for 2,377,510 medium-resolution spectra from LAMOST DR7 is made publicly available online.
This catalog contains the following information: the identifier for the observation spectrum (obsid), the fits file name corresponding to the observation spectrum (filename), coordinate information (ra, dec), the extension name of the spectrum (extname\_blue, extname\_red), the S/N of the spectrum (snr\_blue, snr\_red), effective temperature (Teff[K]), surface gravity (Logg), 15 elemental abundances (XH), $1\sigma$ uncertainty of parameters (X\_err) and recommended flags (flag).
The detailed catalog description information is shown in Table \ref{tab:table3}, and the complete parameter catalog can be downloaded from \href{https://github.com/Chan-0312/RRNet/releases}{https://github.com/Chan-0312/RRNet/releases}.

\section{Summary and Outlook}
\label{sec:Summary}

This paper designed a novel model Residual Recurrent Neural Networks (RRNet) for estimating stellar parameters, computed a LAMOST-RRNet catalog by estimating stellar atmospheric parameters and elemental abundances from LAMOST DR7 medium-resolution spectra using the RRNet.
The RRNet model is trained and tested on reference data from common stars between LAMOST observation spectra and high-precision APOGEE-\textit{Payne} catalog.
With the trained RRNet model, we estimated the stellar atmospheric parameters, chemical abundances and corresponding uncertainties from 2,377,510 medium-resolution spectra in LAMOST DR7.
In case of $\rm{S/N \geq 10}$, the precision of the parameters $T_{\rm eff}$, $\log \, g$, [Fe/H], and [Cu/H] are 88.5K, 0.13 dex, 0.05 dex, and 0.19 dex respectively, while the precision of the other elemental abundances is 0.05 dex $\thicksim$ 0.14 dex.
To verify the performance of the RRNet model, we conducted a series of comparing experiments with other neural network models and other surveys.
The various experiments demonstrate that RRNet has higher accuracy, robustness and has good consistency with other surveys. 
The LAMOST-RRNet catalog for 2.37 million medium-resolution spectra from the LAMOST DR7, the source code, the trained model and the experimental data are released on: \href{https://github.com/Chan-0312/RRNet}{https://github.com/Chan-0312/RRNet}.

In the future, with the increasing of spectral observations and experienced labels, the RRNet can be re-trained to improve its performance and used to estimate stellar atmospheric parameters and elemental abundances from more survey spectra.

\begin{acknowledgments}
This work is supported by the National Natural Science Foundation of China (Grant No. 11973022,  U1811464), and the Natural Science Foundation of Guangdong Province (No. 2020A1515010710).
\end{acknowledgments}

\appendix

\section{RRNet training process}

\begin{figure}[h]
    \centering
    \includegraphics[width=1\linewidth]{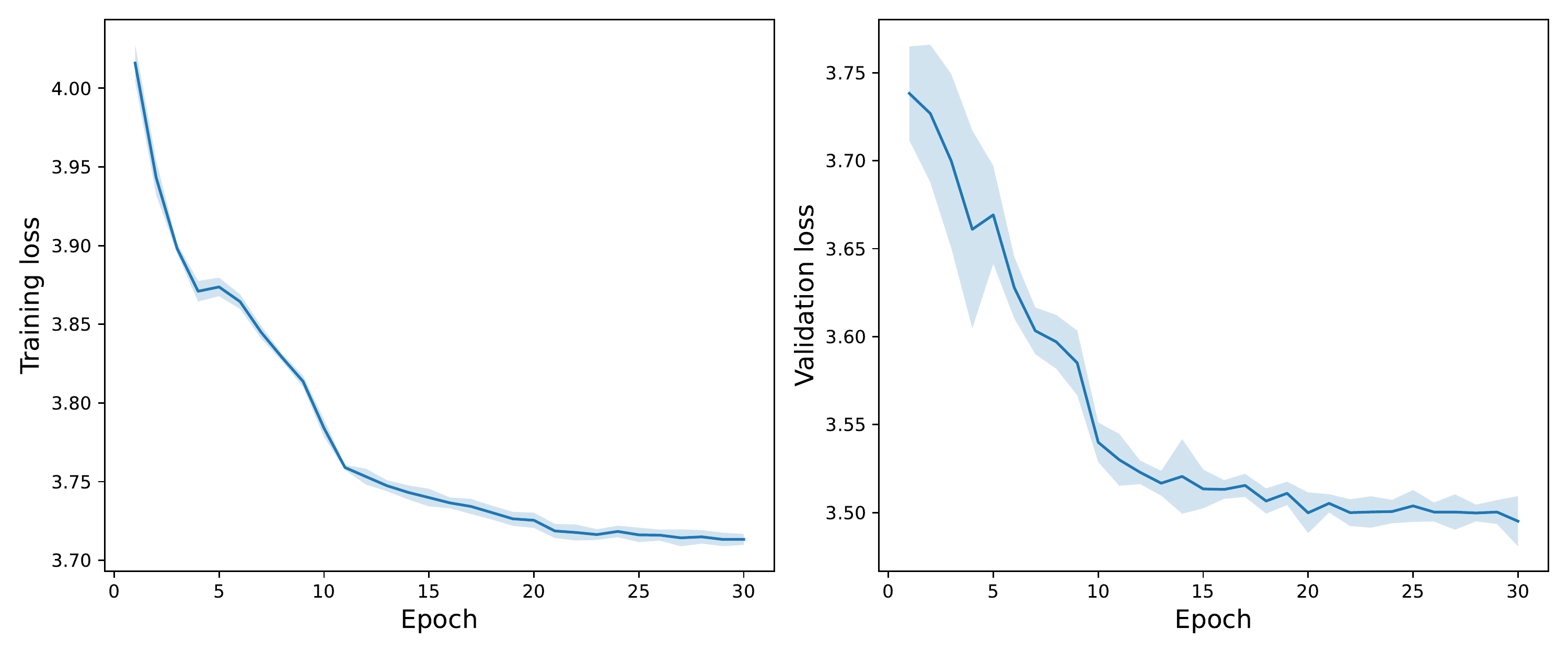}
    \caption{The loss curves of RRNet on the training set and validation set. The solid line indicates the mean of the $M$ model losses ($M = 6$ in this paper), and the shaded area indicates the $1\sigma$ range of the $M$ models.}
    \label{fig:figure10}
\end{figure}

In this section, we will briefly describe the training process of RRNet.
The training and testing of RRNet models are performed on an Nvidia Tesla T4.
We spent about 12 minutes in training each model for 30 epochs with a batch size of 256.
The loss curves of RRNet on the training and validation sets are shown in Figure \ref{fig:figure10}, where the solid line in the figure indicates the average of the losses of $M$ models ($M=6$ in this paper) and the shaded area indicates the $1\sigma$ range of the $M$ models.
The results show that the losses of the training and test sets reach convergence when RRNet is trained for 20 epochs, and the loss curves of all $M$ models are stable and close.
This indicates the reasonableness of the RRNet design.
It should be noted that the overall loss of RRNet in the training process is larger than that in the validation process. This phenomenon results from the Dropout and data augumentation. In test phase, the dropout is disabled. The data augumentation is conducted by generating some novel samples by adding some Gaussian noises onto observed spectrum.

\section{z-score residuals in function of stellar atmospheric parameters}

\begin{figure}[h]
    \centering
    \includegraphics[width=1\linewidth]{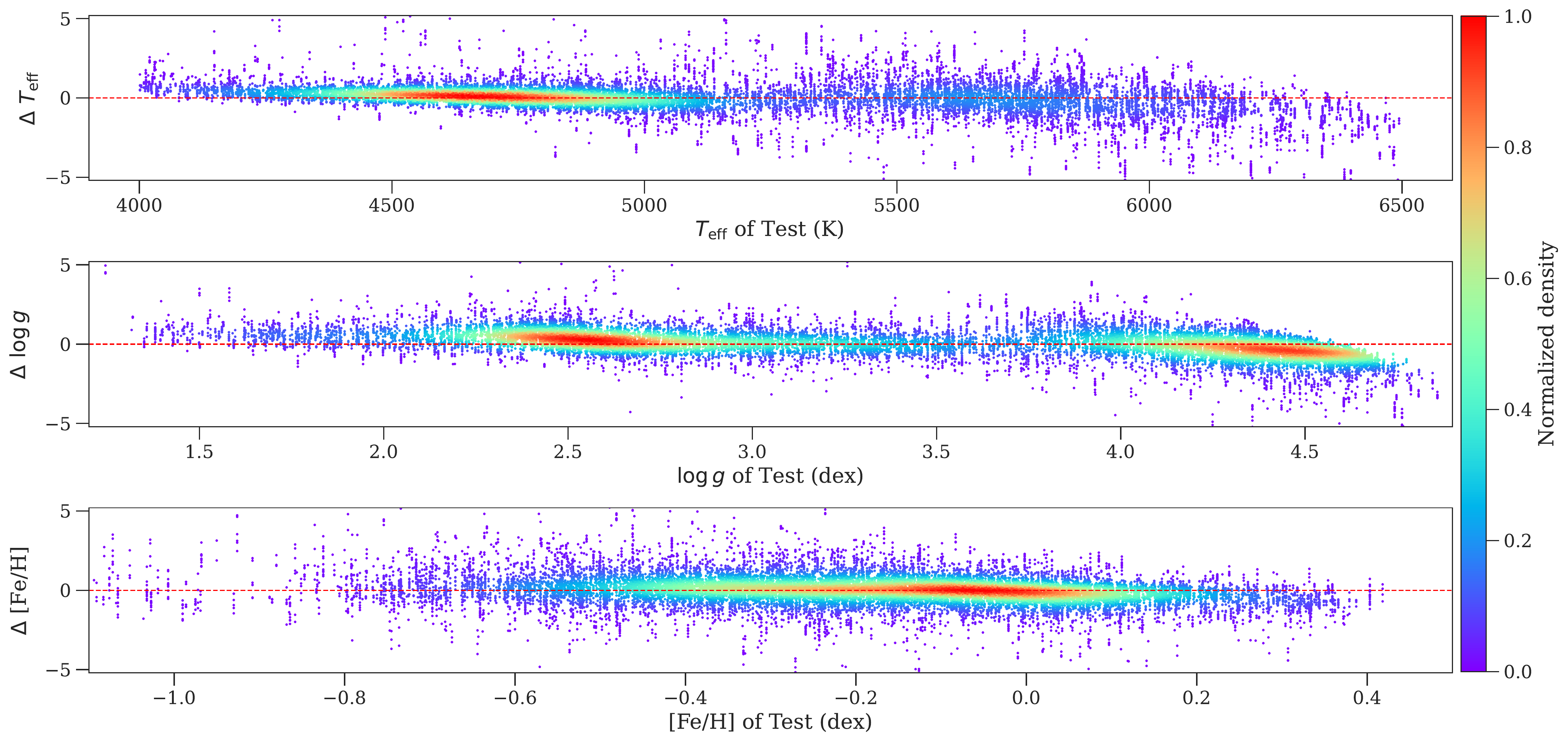}
    \caption{Performance evaluation of RRNet in estimating three stellar atmospheric parameters ($T_{\rm eff}$, $\log \, g$, [Fe/H]).
    In the above figure, the vertical axis presents the distribution of the z-score residuals (difference) between RRNet predictions and APOGEE-\textit{Payne} catalog on test set.
    }
    \label{fig:figure11}
\end{figure}

To further demonstrate the accuracy of RRNet predictions across the range of parameters, Figure \ref{fig:figure11} shows the Z-score residuals (difference) between RRNet predictions and APOGEE-\textit{Payne} catalog for the stellar atmospheric parameters ($T_{\rm eff}$, $\log \, g$, [Fe/H]).
Overall, the RRNet predictions exhibit less bias from the APOGEE-\textit{Payne} catalog.
This is the results that RRNet has Cross-band Belief Enhancement (CBE) capability from the recurrent module. The CBE allows the model to filter out some random noises, and enhances the spectral features based on belief propagation between the observations on different wavelengths.
Furthermore, there is a slight underestimation from RRNet on {$\log \, g$} in case of  {$\log \, g > 4.4 $} (dex). This phenomenon is consistent with the results of \cite{wang2020spcanet}, may result from the intrinsic complexity of the relationship between stellar spectra and {$\log \, g$},  and the scarcity of training examples in this parameter range.

\section{Discussion on wavelength correction}
\label{sec:dis_wavelength_correction}

To simplify the spectral parameter estimation problem based on the available reference set, we perform wavelength correction on the spectrum using the Radial Velocity (RV) provided by the LAMOST catalog.
However, some slight uncertainties may be introduced from the uncertainty of RV.
To explore the effects of RV uncertainty on the RRNet model, we defined four variants of RRNet and conducted some comparison experiments:
1) RRNet-u is trained using uncorrected spectra and its model structure is consistent with RRNet;
2) RRNet-ur is trained using uncorrected spectra similary with the RRNet-u, however, this new variant estimate one more parameter, RV;
3) RRNet-c is the method of this paper, which is trained using the RV-corrected spectra;
4) RRNet-cp is trained from RV-corrected spectra with some augumentation data generated by  RV perturbations.
Table \ref{tab:table4} shows the MAE of the above four model on the validation set.

Overall, the models trained with uncorrected spectra (RRNet-u, RRNet-ur) perform less well than the models trained with corrected spectra (RRNet-c, RRNet-cp).
Therefore, it is necessary to perform wavelength correction on the spectra at the scale of the current training dataset.
Comparing RRNet-u and RRNet-ur, the addition of estimating one more parameter RV reduced the performance of the model.
Comparing RRNet-c and RRNet-cp, data augmentation with an RV perturbation does not significantly improve the performance of the RRNet model.
This may be due to the fact that the recurrent module of RRNet already has the ability to learn features across different bands. Therefore, smaller RV perturbations do not have any significant effects on RRNet.

\begin{deluxetable*}{lcccc}
\tablenum{4}
\tablecaption{Experimental results of RRNet with different pretreatment methods. The model performance is measured using the Mean Absolute Error (MAE, defined in equation \ref{equ:equation4}) and computed from the validation set. \label{tab:table4}}
\tablewidth{0pt}
\tablehead{
\colhead{Labels} & \multicolumn{4}{c}{MAE} \\ 
\cline{2-5} 
\colhead{} & \colhead{RRNet-u}   &  \colhead{RRNet-ur} &  \colhead{RRNet-c}  &  \colhead{RRNet-cp}
}
\decimalcolnumbers
\startdata
$T_{\rm eff}$ & 54.5106    & 57.8453    & 51.8167              & 51.9031                        \\
$\log \, g$        & 0.0872     & 0.0988     & 0.0794               & 0.0792                         \\
{[}Fe/H{]}       & 0.0359     & 0.0385     & 0.0309               & 0.0309                         \\
{[}C/H{]}        & 0.0543     & 0.0575     & 0.0511               & 0.0516                         \\
{[}N/H{]}        & 0.0807     & 0.0836     & 0.0738               & 0.0739                         \\
{[}O/H{]}        & 0.0831     & 0.0831     & 0.0781               & 0.0784                         \\
{[}Mg/H{]}       & 0.0393     & 0.0412     & 0.0357               & 0.0356                         \\
{[}Al/H{]}       & 0.0524     & 0.0529     & 0.0462               & 0.0471                         \\
{[}Si/H{]}       & 0.0375     & 0.0389     & 0.0334               & 0.0332                         \\
{[}S/H{]}         & 0.0603     & 0.0617     & 0.0580               & 0.0587                         \\
{[}K/H{]}         & 0.0925     & 0.0917     & 0.0892               & 0.0892                         \\
{[}Ca/H{]}        & 0.0429     & 0.0443     & 0.0391               & 0.0392                         \\
{[}Ti/H{]}        & 0.0759     & 0.0767     & 0.0732               & 0.0731                         \\
{[}Cr/H{]}        & 0.0685     & 0.0699     & 0.0651               & 0.0648                         \\
{[}Mn/H{]}        & 0.0722     & 0.0739     & 0.0688               & 0.0691                         \\
{[}Ni/H{]}        & 0.0461     & 0.0467     & 0.0411               & 0.0413                         \\
{[}Cu/H{]}       & 0.1383     & 0.1470     & 0.1334               & 0.1341                         \\
\enddata
\tablecomments{RRNet-u, RRNet-ur,
RRNet-c and RRNet-cp are four variants of RRNet by considering the uncertainties of RV, and defined in Appendix \ref{sec:dis_wavelength_correction}.
}
\end{deluxetable*}

\bibliography{main}{}
\bibliographystyle{aasjournal}

\end{CJK}
\end{document}